# Inflation and unemployment in Switzerland: from 1970 to 2050

Oleg Kitov
Department of Economics, University of Oxford
Ivan Kitov
IDG, Russian Academy of Sciences

**Abstract**
An empirical model is presented linking inflation and unemployment rate to the change in the level of labour force in Switzerland. The involved variables are found to be cointegrated and we estimate lagged linear deterministic relationships using the method of cumulative curves, a simplified version of the 1D Boundary Elements Method. The model yields very accurate predictions of the inflation rate on a three year horizon. The results are coherent with the models estimated previously for the US, Japan, France and other developed countries and provide additional validation of our quantitative framework based solely on labour force. Finally, given the importance of inflation forecasts for the Swiss monetary policy, we present a prediction extended into 2050 based on official projections of the labour force level.

Keywords: inflation, unemployment, labour force, forecasting, Switzerland.
JEL classification: E3, E6, J21.

## Introduction

The Swiss National Bank (SNB) introduced a unique monetary policy framework in 2000. The new policy is based on inflation forecasts and utilizes various prediction methodologies (Baltensperger et al., 2007). There are two general approaches as related to conditional and unconditional forecasts. The former approach fixes a constant interest rate over the forecast horizon. The SNB unconditional forecast is calculated under the assumption of endogenous reaction of the nominal interest rate. Both forecasting techniques are based on structural models of the Swiss economy. In addition, unconditional methodology can use non-structural models. The SNB publishes its forecasts on a quarterly basis at a 12-quarter horizon. Lack (2006) presented a comprehensive study of the accuracy of the VAR inflation predictions and found that the best specifications are superior to the naive models of inflation, which became a benchmark model after the paper published by Atkeson and Ohanian (2001).

Lack finds that the most relevant variables for explaining variation in inflation rate are mortgage loans and M3 money supply. It is further reported that real GDP and other real activity indicators only act to the detriment of the forecasting accuracy. Furthermore, Hock and Zimmerman (2005) found that the 3-month LIBOR is superior to other market variables. Nevertheless, all market based forecasts are characterized by large systematic errors. Taking into account critical importance of accurate inflation prediction for the SNB's monetary policy any alternative model providing a more robust and accurate forecast must be carefully considered. Recently, we have revealed a quantitative link between the change rate of labour force and the inseparable pair of price inflation and unemployment in developed countries (Kitov, 2006; Kitov and Kitov, 2010).

Originally, the model was introduced for the United States. The root-mean-square forecasting error (RMSFE) of the rate of price inflation at a 2.5 year horizon was of 0.8% between 1965 and 2004. This RMSFE is superior (at least by a factor of 1.5) to any other model at the same time horizon. Since the change in labour force is not among the predictors used by the SNB, it would be helpful to test the predictive power of this macroeconomic variable for the purposes of devising reliable monetary policy. Therefore, the main purpose of this study is to improve on the current accuracy of description of the past time series of inflation and unemployment using the change in labour force and to predict the future evolution of both variables using labour force projections.

In our previous studies, we have found that the revealed link is represented by a linear and lagged function. Furthermore, since the latter does not include any AR-components the relation between labour force and inflation is a causal one. The corresponding model is completely deterministic with the change in labour force being the only driving force causing all variation in both unemployment and inflation beyond measurement noise. As prescribed by econometric methodology, the models for individual developed countries have been tested for cointegration. This test is mandatory since all involved macroeconomic variables are integrated of order 1, I(1).

A number of standard cointegration tests has been carried out and these tests rejected the null hypothesis of the absence of cointegrating relations (Kitov et al., 2007). Both the Engle-Granger and Johansen approaches have shown the presence of long-term equilibrium relations. Nevertheless, the change in labour force is likely a stochastic process. This subsequently causes inflation and unemployment, as dependent variables of the labour force, to exhibit stochastic properties.

In general, we have been following the approach to macroeconomic modelling of inflation and unemployment adapted in the mainstream economic and econometric research. Regarding the formal representation of inflation and unemployment time series, we follow up the approach of Stock and Watson (2008) who conducted an extensive study of economic and financial variables and indices as predictors of inflation using the Phillips curve framework. For some reason the set of 200 predictors does not include the change in labour force.

In this paper, we aim to correct for this drawback and extend the predictors' set in order to conduct a similar statistical investigation for Switzerland. However, we completely ignore any autoregressive (AR) components in the underlying time series. Thus, our model does not belong to the class of Autoregressive Integrated Moving Average (ARIMA) or Unobserved Component - Stochastic Volatility (UC-SV) models. It is rather a model with degenerate stochastic components, except those associated with measurement noise.

The Phillips curve in Switzerland was modelled by Gerlach (2007), who addressed the question of how useful money growth is for explaining future price developments. The SNB used money growth targets until 1999. When the new policy framework focusing on inflation forecasts was introduced money growth has been used as a major indicator variable. To model the money growth influence on the Phillips curve the trend money growth was incorporated. This additional variable explained around 15% of variability in the original inflation time series. Although the author reports a better predictive power of the extended model, in fact it does not result in any marginal improvements relatively to the other predictors from the set proposed by Stock and Watson (2008).

As expected from our previous experience for a number of developed countries, the change in labour force in Switzerland is characterized by a much better predictive power and inflation can be accurately forecasted even in the absence of its lagged or expected values, as would have been prescribed the Classical New Keynesian framework. For the rate of unemployment, the predictive power of labour force is also rather high. However, the rate of unemployment has been developing in a narrow channel with little variation, except for the leap from 1.7% in 1991 to 3.8% in 1993. Stable processes are harder to model due to extremely low resolution in the data (i.e. naive model or historical mean usually outperforms structural models).

The reminder of the paper is organized in five sections. Section 2 introduces the model. In many countries, the U.S. and Japan among others, the generalized link between labour force and the two dependent variables can be split into two independent relationships, with no apparent explicit link between inflation and unemployment. In France, only the generalized model provided an adequate description of the evolution of both dependent variables since the 1960s (Kitov, 2007). In Section 2, we also present and statistically characterize various estimates of inflation, unemployment and labour force in Switzerland.

Section 3 discusses the Phillips curve and reveals the break in relevant time series around 1994. This year introduces a structural break in some relationships estimated in this study. In Section 4, the linear link between labour force and unemployment is modelled using annual measurements of both variables. Section 5 is devoted to the link between the rate of inflation and labour force. Instead of poorly constrained LSQ methods we apply a simplified version of the 1-D boundary elements method, namely the cumulative curves. Finally, Section 6 presents the generalized link between inflation, unemployment and labour force. In Conclusion, we present several long-term (through 2050) forecasts of inflation in Switzerland using labour force projections provided by the Swiss Federal Statistics Office (2011).

1. **Model and data**

As originally defined in Kitov (2006), inflation and unemployment are linear and potentially lagged functions of the change rate of labour force:

$$\pi(t) = A_1 dLF(t-t_1)/LF(t-t_1) + A_2 \quad (1)$$
$$UE(t) = B_1 dLF(t-t_2)/LF(t-t_2) + B_2 \quad (2)$$

where $\pi(t)$ is the rate of price inflation at time t, as represented by some standard measure such as the GDP deflator (DGDP) or CPI; $UE(t)$ is the rate of unemployment at time t, which can be also represented by various measures; $LF(t)$ is the level of labour force at time t; $t_1$ and $t_2$ are potential time lags between inflation, unemployment, and labour force, respectively; $A_1$, $B_1$, $A_2$, and $B_2$ are country specific coefficients, which have to be determined empirically in the calibration procedures. The latter coefficients may vary through time for a given country, as induced by numerous revisions to definitions and measurement methodologies of the studied variables, i.e. by genuine variations in measurement units. Note that in (1) and (2) the term dt in the relative rate of change of labour force is omitted as the resulting empirical model is based on annual readings and hence dt = 1.

Linear relationships (1) and (2) define inflation and unemployment separately. These variables are two indivisible manifestations or consequences of a unique process, however. The process is the growth in labour force which is accommodated in developed economies (we do not include developing and emergent economies in this analysis) through two channels. First channel is the increase in employment and corresponding change in personal income distribution (PID). All persons obtaining new paid jobs or their equivalents presumably change their incomes to some higher levels. There is an ultimate empirical fact, however, that PID in the USA does not change with time in relative terms, i.e. when normalized to the total population and total income. The increasing number of people at higher income levels, as related to the new paid jobs, leads to a certain disturbance in the PID. This over-concentration (or "over-pressure") of population in some income bins above its "neutral" long-term value must be compensated by such an extension in the corresponding income scale, which returns the PID to its original density. Related stretching of the income scale is the core driving force of price inflation, i.e. the US economy needs exactly the amount of money, extra to that related to real GDP growth, to pull back the PID to its fixed shape. The mechanism responsible for the compensation and the income scale stretching, should have some positive relaxation time, which effectively separates in time the source of inflation, i.e. the labour force change, and the reaction, i.e. the inflation.

The second channel is related to those persons in the labour force who failed to obtain a new paid job. These people do not leave the labour force but join unemployment. Supposedly, they do not change the PID because they do not change their incomes. Therefore, total labour force change equals unemployment change plus employment change, the latter process expressed through lagged inflation. In the case of a "natural" behaviour of the economic system, which is defined as a stable balance of socio-economic forces in the society, the partition of labour force growth between unemployment and inflation is retained through time and the linear relationships hold separately. There is always a possibility, however, to fix one of the two dependent variables. Central banks are definitely able to influence inflation rate by monetary means, i.e. to force money supply to change relative to its natural demand. To account for this effect one should use a generalized relationship as represented by the sum of (1) and (2):

$$\pi(t) + UE(t) = A_1 dLF(t-t_1)/LF(t-t_1) + B_1 dLF(t-t_2)/LF(t-t_2) + A_2 + B_2 \qquad (3)$$

Equation (3) balances the change in labour force to inflation and unemployment, the latter two variables potentially have different time lags behind the labour force change. Effectively, when $t_1 \neq 0$ or/and $t_2 \neq 0$, one should not link inflation and unemployment for the same year. One can rewrite (3) in a form similar to that of the classical Phillips curve, but without the autoregressive terms:

$$\pi(t) = C_1 dLF(t-t_1)/LF(t-t_1) + C_2 UE(t+t_2-t_1) + C_3 \qquad (4)$$

where coefficients $C_1$, $C_2$, and $C_3$ are naive summations of coefficients from the above equations. Nevertheless, these will be determined using separate calibration. The rationale behind the superiority of the empirical estimation is the presence of high measurement noise in all original time series. In some places, (4) can provide a more effective destructive interference of such noise than does (3). Consequently, the coefficients providing the best fit for (3) and (4), whatever method is used, may be different.

The principal source of information is the OECD database (http://www.oecd.org/) which provides comprehensive data sets on labour force, unemployment, GDP deflator, and CPI inflation. We also used the estimates reported by the U.S. Bureau of Labor Statistic (http://www.bls.gov) for corroboration of the data on CPI, unemployment and labour force. As a rule, readings associated with the same variable but obtained from different sources do not coincide. This is due to different approaches and definitions applied by corresponding agencies. This diversity of definitions is accompanied by a degree of uncertainty related to the methodology of measurements.

This uncertainty cannot be directly estimated but has an explicit effect on reliability of empirical relationships (Sims, 2009). At first, we introduce various estimates of all involved variables. There are two time series for the rate of inflation and unemployment. The level of labour force is represented by annual and quarterly readings. Figures 1 displays the evolution of two principal measures of price inflation in Switzerland: the GDP deflator, DGDP, and consumer price index, CPI. Both variables are published by the OECD. As it has been already discussed, we consider the DGDP time series a better representative of price inflation in a given country since it comprises all prices in an economy. The overall consumer price index is fully included in the DGDP, and thus, its behaviour represents only a part of the economy. Since labour force and unemployment do characterize the entire economy it is methodically better to use the DGDP for any quantitative modelling.

Figure 1 shows that the overall difference between the CPI and DGDP is not prominent but there are short periods of very large discrepancy: in the beginning of the 1970s, from 1985 to 1989, from 1991 to 1993, and around 2005. The DGDP time series starts only in 1971 whereas CPI is available since the 1950s.

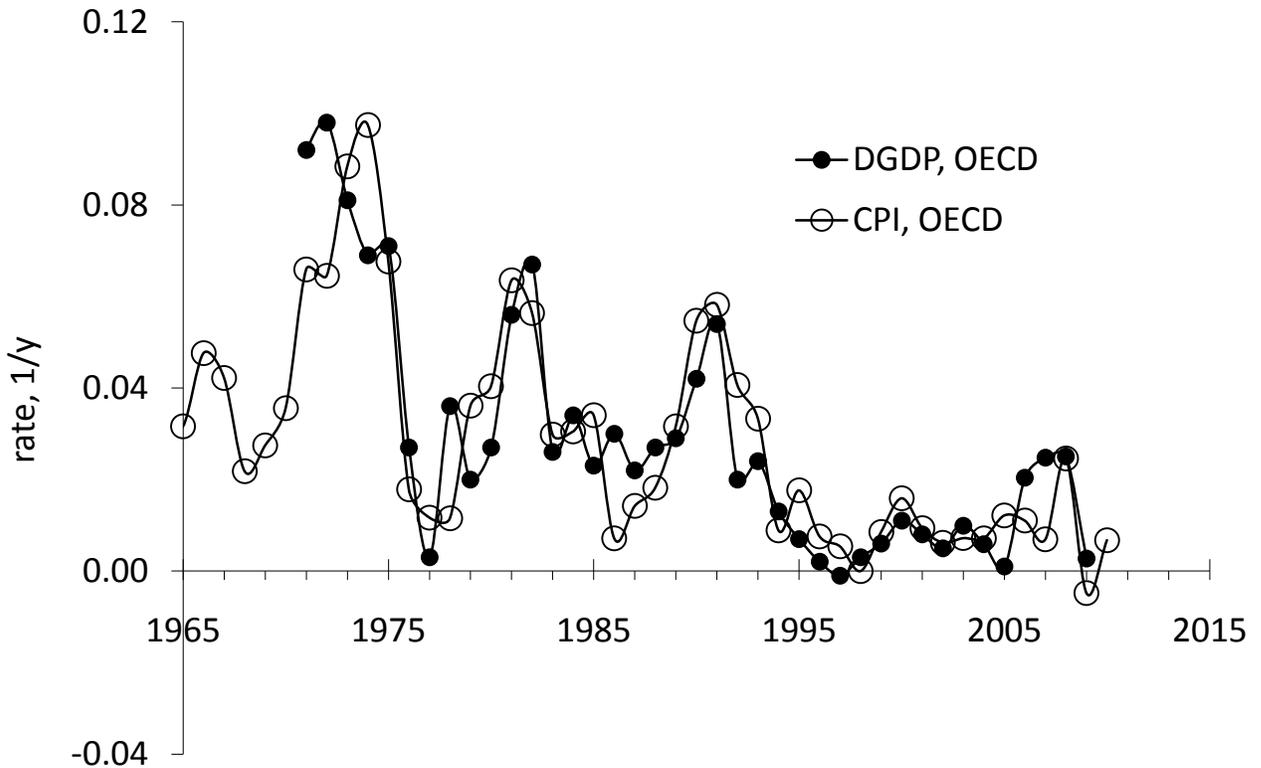

**Figure 1: Two definitions of the rate of price inflation in Switzerland: the GDP deflator (available from 1971 to 2009) and CPI inflation (displayed from 1965 to 2010) as published by to the OECD.**

The rate of inflation fell to $0.01y^{-1}$ in 1994 and has been oscillating around this level since. Between 1971 and 1993, inflation was fluctuating with decreasing amplitude with apparent trend persistent periods of around 8 years on average. There are three sharp peaks in 1974, 1982 and 1990. Smaller peaks were observed in 1999 and 2008. Such behaviour is different from that observed in most developed countries including those we analyzed previously. It is thus a challenge for our concept and an opportunity to validate the model.

For further econometrical analysis of the link with the labour force, one needs to estimate basic properties of both time series. The order of integration can be determined using unit root tests applied to the original series and their progressive differences. As a rule, the rate of inflation is a I(1) variable in developed countries. Switzerland is not an exception, as Table 1 clearly demonstrates. In particular, we report the results of the following tests: the augmented Dickey-Fuller (ADF), the DF-GSL, and the Phillips-Perron test. The DGDP series consists of only 39 readings and the results might not be reliable. Nevertheless, all tests do not reject the null of the presence of a unit root. For the CPI series with 50 readings the principal result is the same. Both _first differences (dDGDP and dCPI) are characterized by the absence of unit roots, and thus, the original time series are integrated of order 1. For the period between 1971 and 2009, the RMS values for the DGDP and CPI are 0.262 and 0.257, respectively. For the naive prediction of inflation at a one-year horizon between 1972 and 2009, the RMS values (e.g. DGDP(t) - DGDP(t - 1)) are 0.1631 and 0.1630, respectively. We accept these latter values as the benchmark accuracy for any prediction.

Figure 2 depicts two estimates of the rate of unemployment as reported by the OECD. The first one is adherent to a standard measurement technique, UE, and the other is measured according to a new definition and is referred to as the harmonized rate of unemployment, HUE, as described by the OECD (2011). The difference between the two curves is almost negligible and thus without any potential loss of accuracy we employ the standard rate, UE, for the rest of the paper. The latter has been measured since 1975.

**Table 1: Unit root tests of the original time series and their first differences.**

|            | DF              | DF-GLS (lag 2)   | PP z($\rho$)      | z(t)             |
|------------|-----------------|------------------|-------------------|------------------|
| DGDP       | -2.76 (-3.66)[1]| -2.82 (-3.77)    | -8.26 (-18.08)    | -2.68 (-3.66)    |
| dDGDP      | -6.73* (-3.67)  | -4.43* (-3.77)   | -40.05* (-18.01)  | -6.79* (-3.67)   |
| CPI        | -2.45 (-3.59)   | -3.66 (-3.77)    | -13.88 (-18.83)   | -2.68 (-3.59)    |
| dCPI       | -5.86* (-3.60)  | -3.67 (-3.77)    | -38.55* (-18.70)  | -5.80* (-3.60)   |
| UE         | -0.83 (-3.69)   | -2.42 (-3.77)    | -2.20 (-17.18)    | -1.00 (-3.69)    |
| dUE        | -3.37 (-3.67)   | -3.45 (-3.77)    | -17.98* (-17.74)  | -3.32 (-3.69)    |
| dLF/FL     | -4.53* (-3.59)  | -2.71 (-3.77)    | -28.00* (-18.76)  | -4.50* (-3.59)   |
| d(dLF/LF)  | -9.18 *(-3.60)  | -4.81* (-3.77)   | -55.27* (-18.70)  | -9.73* (-3.60)   |

[1] 1% critical value; * - the null of a unit root is rejected.

The shape of the unemployment curve is rather peculiar and can be roughly represented by two quasi-constant segments: before 1991 and after 1993. The same overall structure is observed in Austria with the leap from 1.7% in 1981 to 4.2% in 1983. For Austria, this step was artificially introduced by the change in definition of the unemployment variable. Hence, one might expect the same cause behind the step in the rate of Swiss unemployment. Such artificial steps can be modelled by relevant changes in all coefficients in (2) and do not represent actual structural breaks in the original time series or the unemployment volatility associated with underground sector (Gaetano, 2010). Therefore, the concept of unemployment has many aspects to elaborate on (Deutsch et al., 2007).

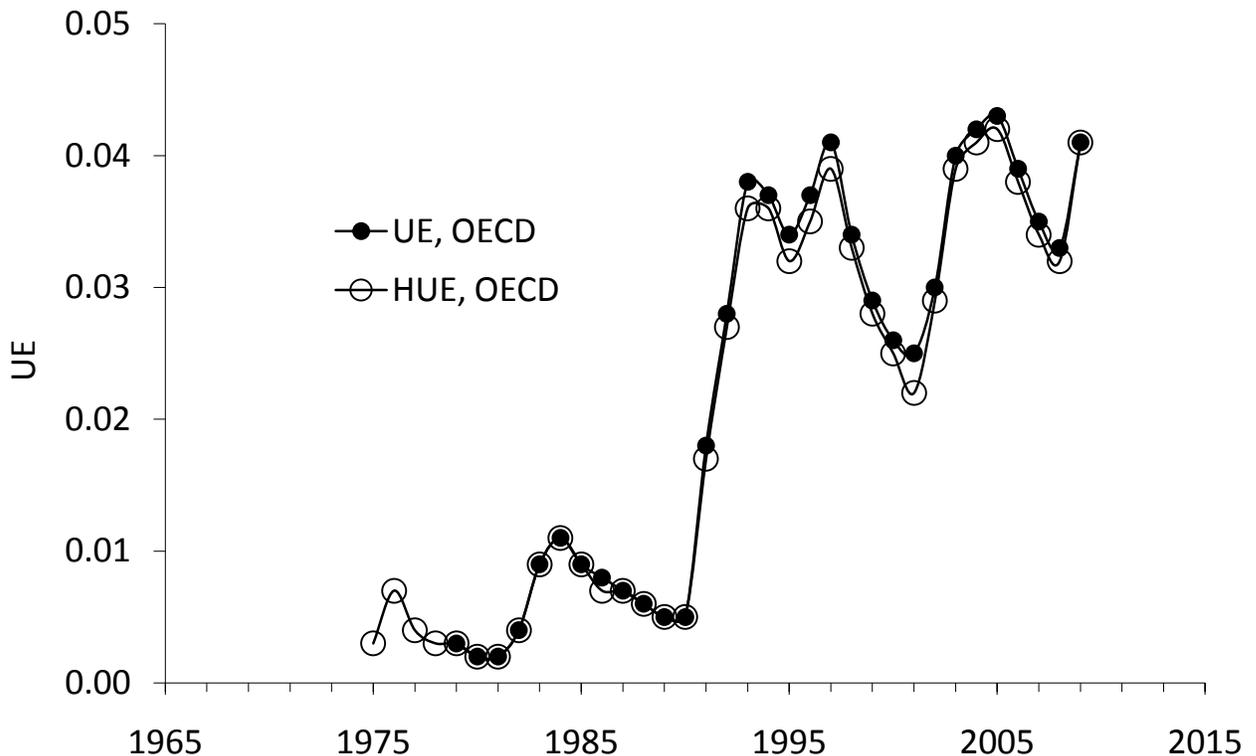

**Figure 2: The rate of unemployment in Switzerland according to two OECD definitions.**

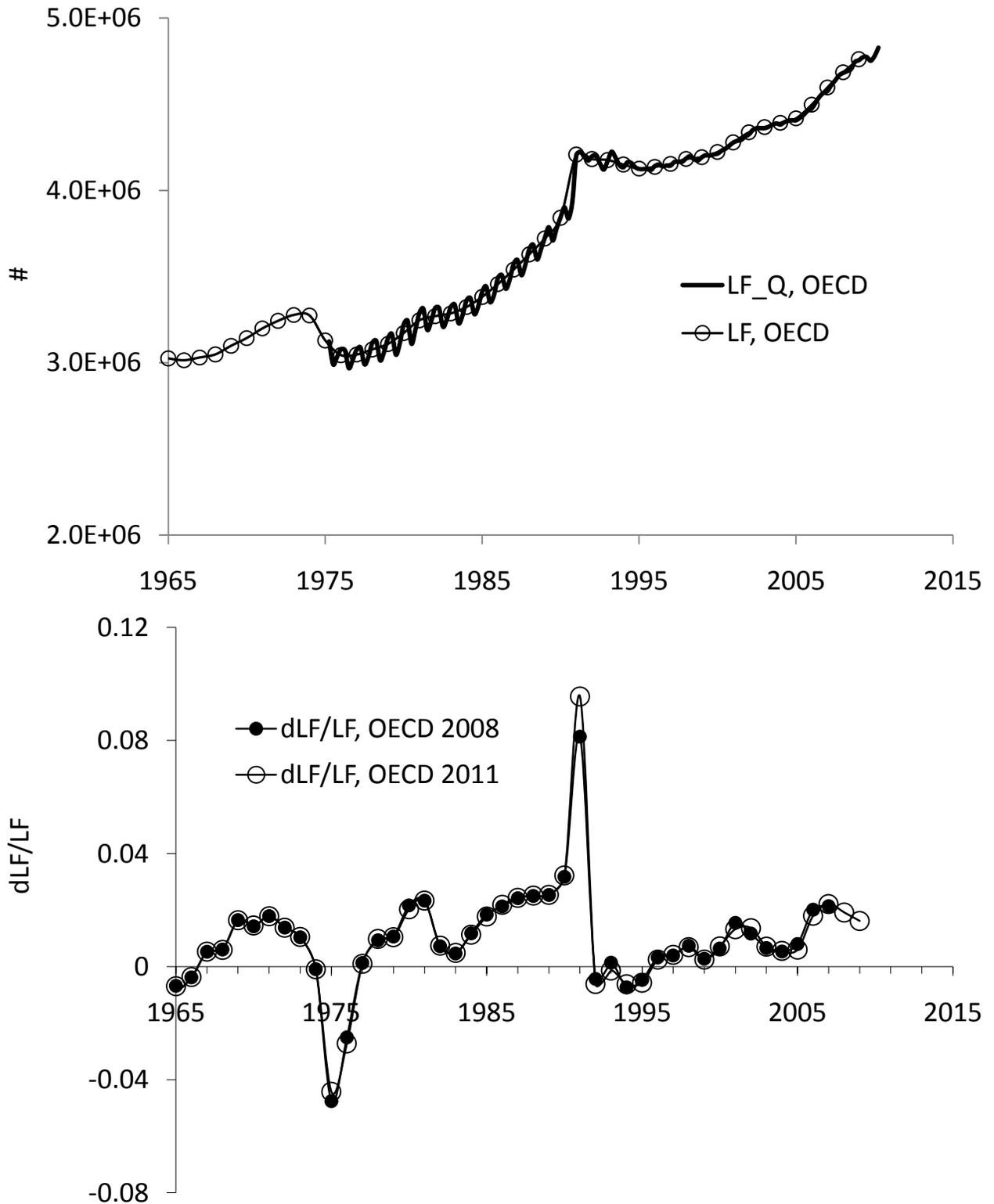

**Figure 3:** *Upper panel*: The level of labour force: annual, LF, and quarterly, LF$_Q$, estimates. *Lower panel*: The rate of annual labour force change in Switzerland. For the annual change, two versions of labour force are presented: 2008 and 2011 vintages.

In Switzerland, the highest rate of measured unemployment was at the level of 4.3% in 2005, and the lowermost was registered in 1981 at 0.2%. This extremely low rate of unemployment was likely related to the definition and/or the measurement problems. In any case, the change in unemployment is so insignificant that one would not expect any quantitative modelling for the period after 1975 to produce any statistically robust results. Despite the step between 1991 and 1994, Table 1 demonstrates that the UE series is likely integrated of order 1, with the first difference series having unit roots or eigenvalues very close to 1.0. Yilanci (2008) found no unit roots in the monthly estimates of the unemployment rate in Switzerland between 1993 and 2008: the dramatic rise between 1991 and 1993 was safely ignored.

The level of labour force, LF, has been measured since the 1950s. Figure 3 depicts the annual and quarterly (since 1975) estimates of the LF and the rate of change in the annual levels, dLF/LF. Apparently, the labour force series has two, at least in part of artificial nature, breaks: one in 1974 of unknown nature and one in 1992, as the OECD (2008) informs:

> ***Series Breaks****. From 1998, data are adjusted in line with the 2000 census. Prior to 1991, data refer only to persons who are gainfully employed at least six hours per week.*

In the lower panel of Figure 3, two versions of labour force change are presented: the 2008 and 2011 vintages. The difference in the change rate is minor over the entire period, except in 1991. This difference demonstrates the uncertainty in the labour force estimates conducted during the labour force surveys. Table 1, indicates that the annual dLF/LF is rather an I(1) process with unit roots or eigenvalues very close to 1.0. The first difference, d(dLF/LF), is an I(0) process with all tests rejecting the null hypothesis of a unit root. Since linear links between three I(1) processes will be estimated later, it is necessary to carry out specific tests for cointegration. Otherwise any statistical estimates can be biased by stochastic trends in these time series. As mentioned above, small samples do not guarantee a higher reliability of statistical tests and we use quarterly estimates of inflation and labour force for unit root tests.

Table 2 lists some results of several unit root tests of the quarterly DGDP (119 readings at Q/Q basis) and dLF/LF (139 readings) series and their first differences. Monthly and quarterly estimates of the UE are also available; however, we do not expect the model of the UE to be a reliable one because of low variability in the data, and thus, low dynamic resolution in this time series. As for the annual estimates, both quarterly time series demonstrate the presence of unit roots in level and the absence of unit roots in the first differences. Therefore, both time series are generated by I(1) processes.

**Table 2: Unit root tests of the quarterly estimates of the DGDP and dLF/LF and their first differences**

|  | DF | DF-GLS (lag 4) | PP z($\rho$) | PP z(t) |
|---|---|---|---|---|
| **DGDP** | -1.24 (-3.50) | -1.29 (-3.56) | -7.49 (-19.86) | -1.17 (-3.50) |
| **dDGDP** | -8.46* (-3.50) | -4.99* (-3.56) | -89.61* (-19.86) | -8.48* (-3.50) |
| **dLF/FL** | -2.09 (-3.497) | -1.93 (-3.53) | -18.28 (-19.93) | -2.66 (-3.50) |
| **d(dLF/LF)** | -9.32 *(-3.498) | -4.69* (-3.53) | -100.4* (-19.92) | -9.20* (-3.50) |

2. **The Phillips curve**

The Phillips curve is a prominent statistical link between price inflation and unemployment. As a rule, it is represented by a scatter plot. In Figure 4, we illustrate the Phillips curve in Switzerland as time series of the rate of unemployment against the rate of CPI inflation reduced to the unemployment series by a linear relationship. The period between 1975 and 2009 has to be separated into two segments according to the artificial step around 1993:

UE(t) = -0.2CPI(t+1) + 0.014; t < 1994
UE(t) = -1.0CPI(t+1) + 0.04;  t > 1993                                    (5)

Both slopes in (5) are negative. For the period after 1993, the slope of -1.0 means that any increase in the overall price at a rate above $0.04y^{-1}$ would reduce the rate of unemployment to zero. When the rate of price inflation is zero, the rate of unemployment is constant at the level of 4%. Before 1993, both coefficients are much smaller, i.e. the change in unemployment was much less sensitive to the change in price. Obviously, the main reason for that was the past definition of an unemployed person, which included less people. The change in the coefficients is of purely artificial change in measurement units.

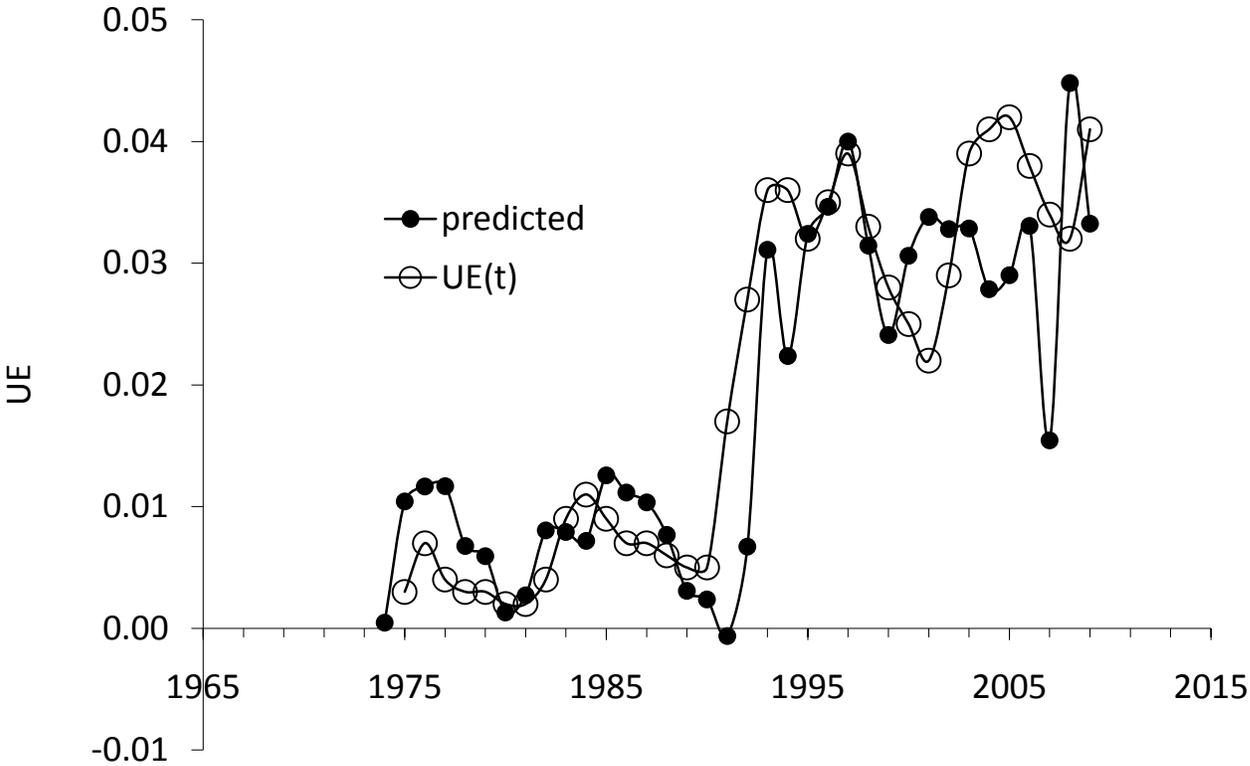

**Figure 4: The Phillips curve in Switzerland between 1975 and 2009.**

According to (5), the CPI lags behind the rate of unemployment by one year. This is a prominent feature of unemployment in Switzerland and completely corresponds to the conventional understanding of the Phillips curve - any change in the rate of unemployment causes a proportional change in the rate of price inflation. This is not a common feature in developed economics. In the U.S. and Italy, the change in unemployment lags behind inflation. In Japan, Austria and Australia, there is no lag between these macroeconomic variables. In any case, one can use this one-year lag and predict the CPI from the UE.

All in all, one can conclude that the Phillips curve does exist in Switzerland for the entire period between 1975 and 2009 as a single relationship between unemployment and lagged inflation, when the change in measurement units is compensated by the coefficients change in (5). Here, statistical assessments of the Phillips curve are skipped since the overall relationship contains an artificial major step which would bias any estimates. Note that the coefficients in (5) were estimated by visual fit between the observed and predicted curves, and thus, are not accurate.

### 3. Unemployment and labour force

After the Phillips curve, our approach dictates that we model the dependence of unemployment on the change in labour force. We proceed by replacing the rate of inflation with the rate of labour force change in (5) and then estimate new coefficients and lags. It has been empirically revealed and statistically tested that the rate of unemployment in developed countries is a linear lagged function of the change in labour force. In theory, there should be no reason why a similar link would not exist for Switzerland.

As expected, the same relationship is valid for Switzerland. The estimation method is as before - the trail-and-error one. For the annual readings in Figure 5, we do not use the cumulative curve approach and fit only peaks and troughs. The best-visual-fit equations for the period before and after 1991 (the start of the step in the unemployment curve) are as follows:

$$UE(t) = -0.1dLF(t)/LF(t) + 0.007; \ t < 1992$$
$$UE(t) = -0.5dLF(t)/LF(t) + 0.04; \ \ t > 1991 \tag{6}$$

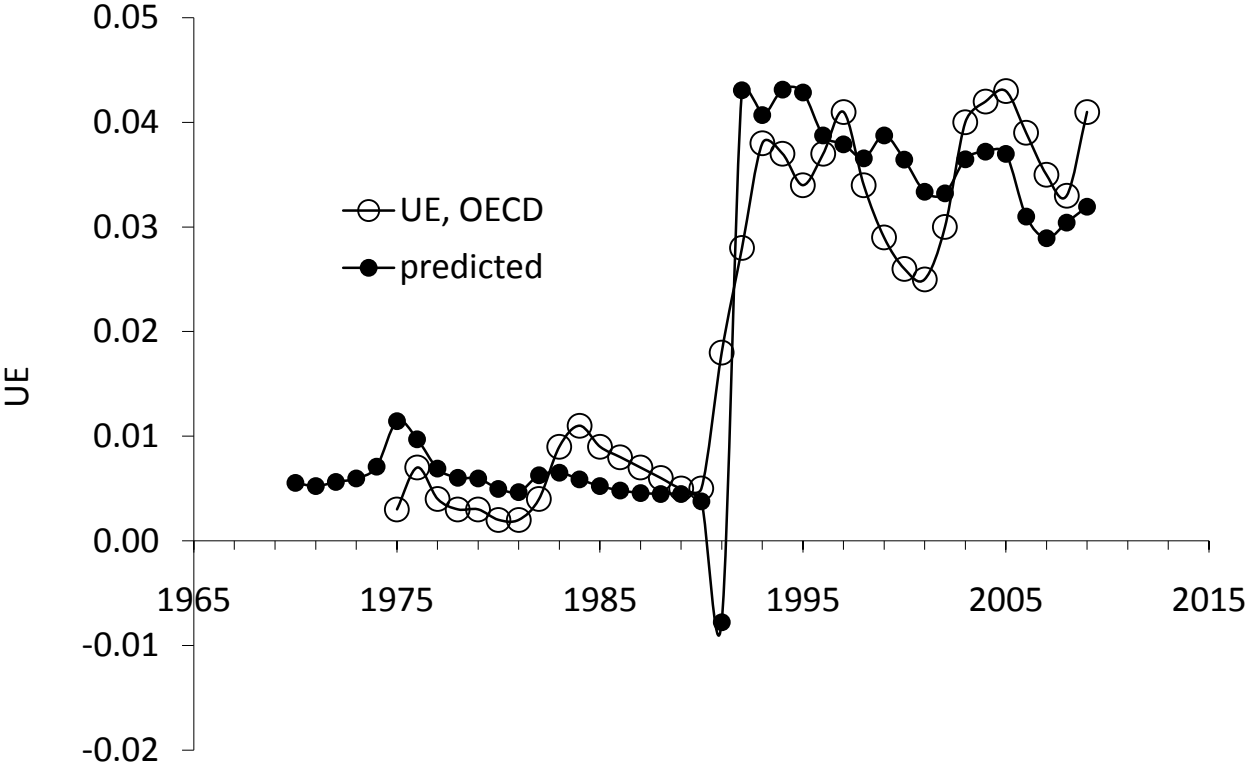

**Figure 5: The measured rate of unemployment in Switzerland and that predicted from the change rate of labour force.**

There is no time delay between the UE and dLF/LF. Therefore, the break in the Phillips curve observed in 1993 is shifted by one year in the past. This break corresponds to the major artificial step in the LF series. Thus, both time series share the same revision to definitions. Apparently, the slope in (6) is negative. Therefore, any increase in the level of labour force is reflected in a proportional and simultaneous fall in the rate of unemployment. This is a fortunate causality direction - more jobs are equivalent to less unemployed. When the level of labour force does not change over time, the rate of unemployment stays at a low mark of 4%.

It is worth noting that (6) implies a nonlinear dependence on the rate of participation in labour force which is a common complication for economic sciences (Ungureanu and Matei, 2007). For a given absolute change in the level of labour force, say 100,000 per year, the reaction of unemployment with be different for the rate of participation 50% (in 1980) and 61% (in 2009). The higher is the participation rate the lower is the change rate, dLF/LF, and thus, the change in the rate of unemployment. It will be a difficult task to retain the rate of unemployment at the current low level - it is likely that the rate of participation is approaching the peak level and will start to decline in the near future.

### 4. Inflation and labour force

The existence of a deterministic link between labour force and price inflation has been found for many countries. Here we are following the estimation procedure based on cumulative curves, which was successfully employed in Kitov and Kitov (2010). We start with the annual readings of CPI inflation as reported by the OECD. According to the spike in the dLF/LF (associated with the change in definition) in 1991 we have divided the period before and after 1991. An additional break was observed in 1979. Thus, we separate the overall relation into three segments. By fitting the cumulative curves, we have obtained the following empirical models:

$$CPI(t) = 1.9dLF(t-1)/LF(t-1) + 0.053; t < 1981$$
$$CPI(t) = 1.3dLF(t-1)/LF(t-1) + 0.008; 1992 > t > 1980$$
$$CPI(t) = 0.5dLF(t-1)/LF(t-1) + 0.006; t > 1991 \qquad (7)$$

All slopes in (7) are positive and the lag is one year. Thus, any positive increment in labour force in Switzerland causes a proportional overall price increase in one year. Since 1991, the slope is +0.5 and 1% change in the LF has to result in 0.5% change in the rate of inflation. When the level of labour force is stable, the rate of inflation is only +0.6% per year. One can interpret this free term as a low level "intrinsic inflation persistence" in line with the conclusion made for Switzerland by Benati (2009) and Elmer and Maag (2009) and speculate on the influence of the explicit inflation targeting on the long-term level of inflation (Libich, 2006). Before 1991, the slope is larger reflecting relatively smaller absolute changes in the labour force under the previous definitions.

Figure 6 displays the observed CPI inflation curve and that predicted according to (7). Both, the annual and cumulative curves are in agreement. For the period between 1971 and 2009, the goodness of fit is very high: $R^2_{ann} = 0.64$ and $R^2_{cum} = 0.997$, respectively. If to consider that (7) does not use autoregressive properties of inflation, which usually bring between 70% and 90% of the explanatory power in the NKPC and other type macroeconomic models, this link is almost a deterministic one. Later in this Section we show that the series of inflation and the change in labour force are cointegrated.

In Figure 7, we have smoothed the predicted and observed series with MA(3) since the original dLF/LF and CPI series are volatile due to insufficient resolution. This smoothing is not to remove stochastic component but to increase the baseline for the LF and CPI estimates. The overall improvement is striking, with all major peaks and troughs well predicted in time and amplitude. Thus, stochastic components in both time series are independent and likely related to measurements noise which is easily suppressed by smoothing. For the annual estimates, one has to admit that any economic research desperately needs accurate measurements to reveal robust statistical ties.

Figure 8 then illustrates the benefits of the cumulative curve approach. The relative modelling error, i.e. the error divided by the observed cumulative inflation, decreases over time. Despite the annual levels of price and labour force are not measured more accurately with time the overall change in the level is measured with increasing precision. As a consequence, the observed and predicted cumulative curves, i.e. the overall changes in price and labour force, do converge. They become indistinguishable, and so there does exists a strong a deterministic link between the overall price level and the level of labour force.

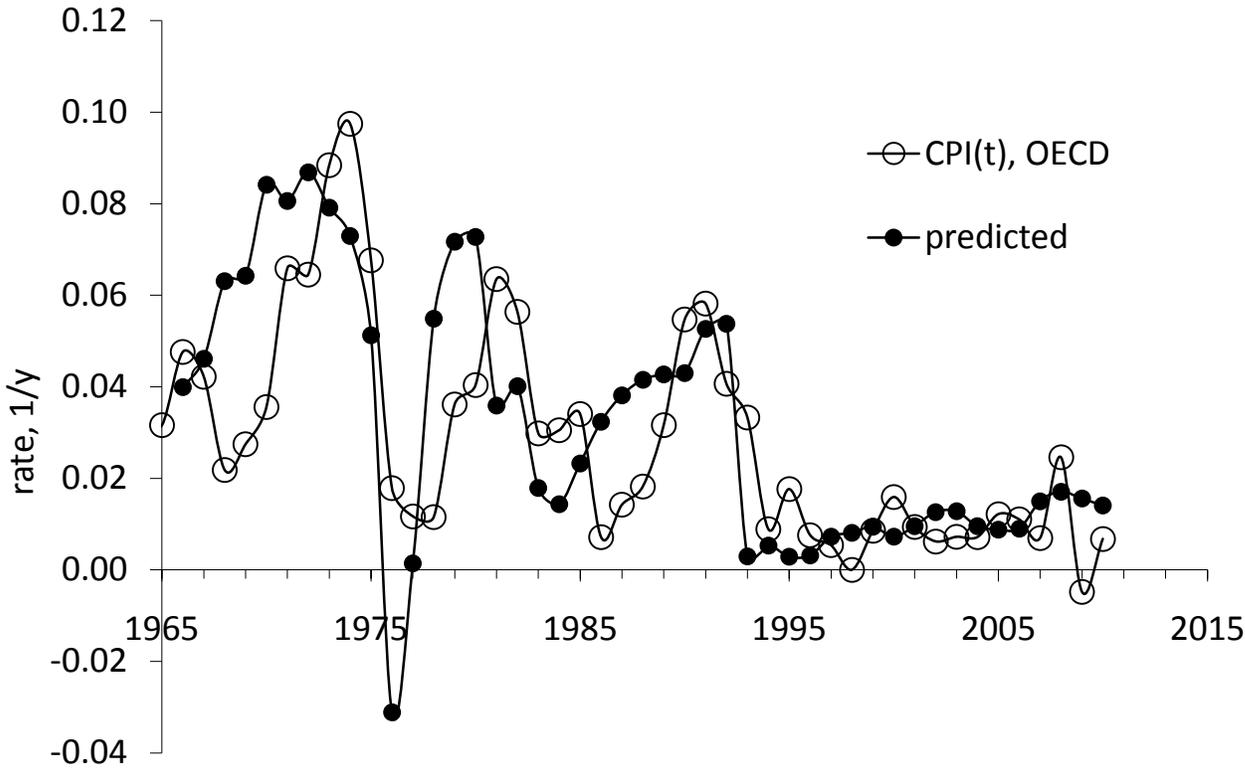

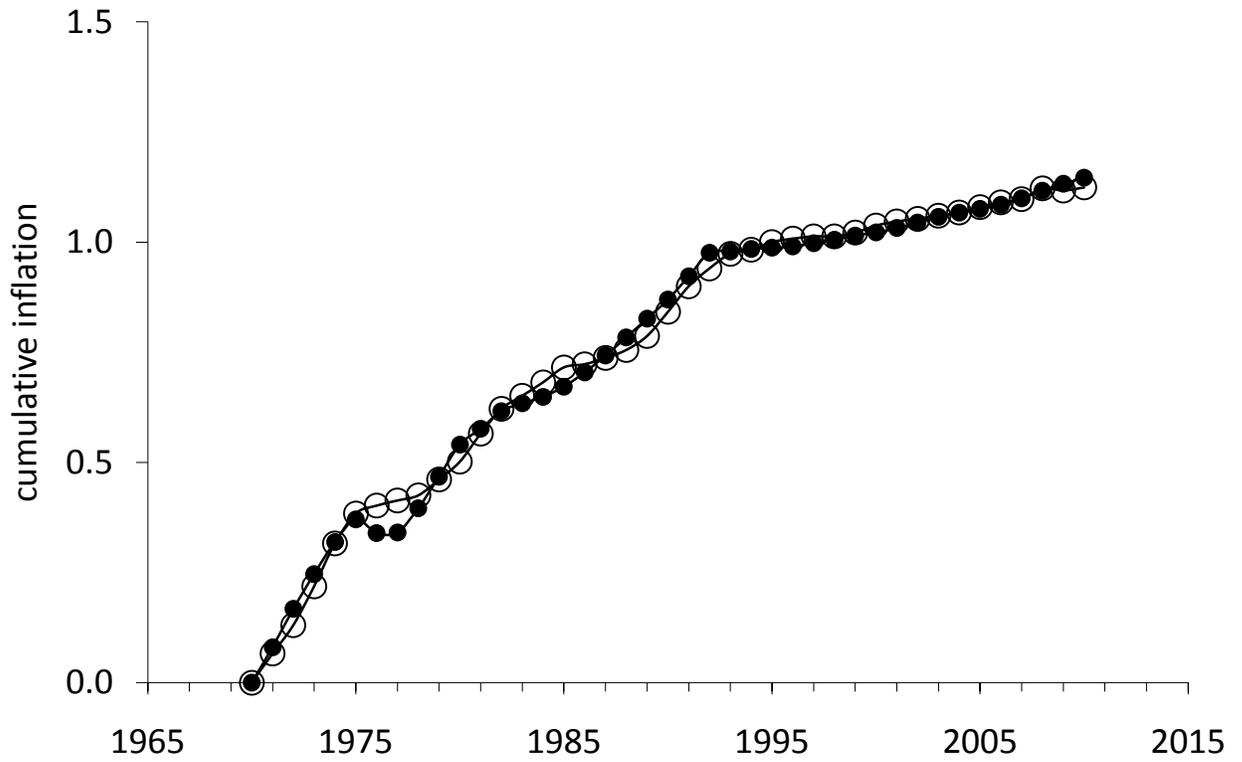

**Figure 6:** *Upper panel*: Annual readings of the observed and modelled rate of CPI inflation. According to the changes in labour force definition in 1979 and 1991, the predicted curve is a piecewise linear and lagged function. *Lower panel*: Cumulative curves of those in the upper panel since 1971.

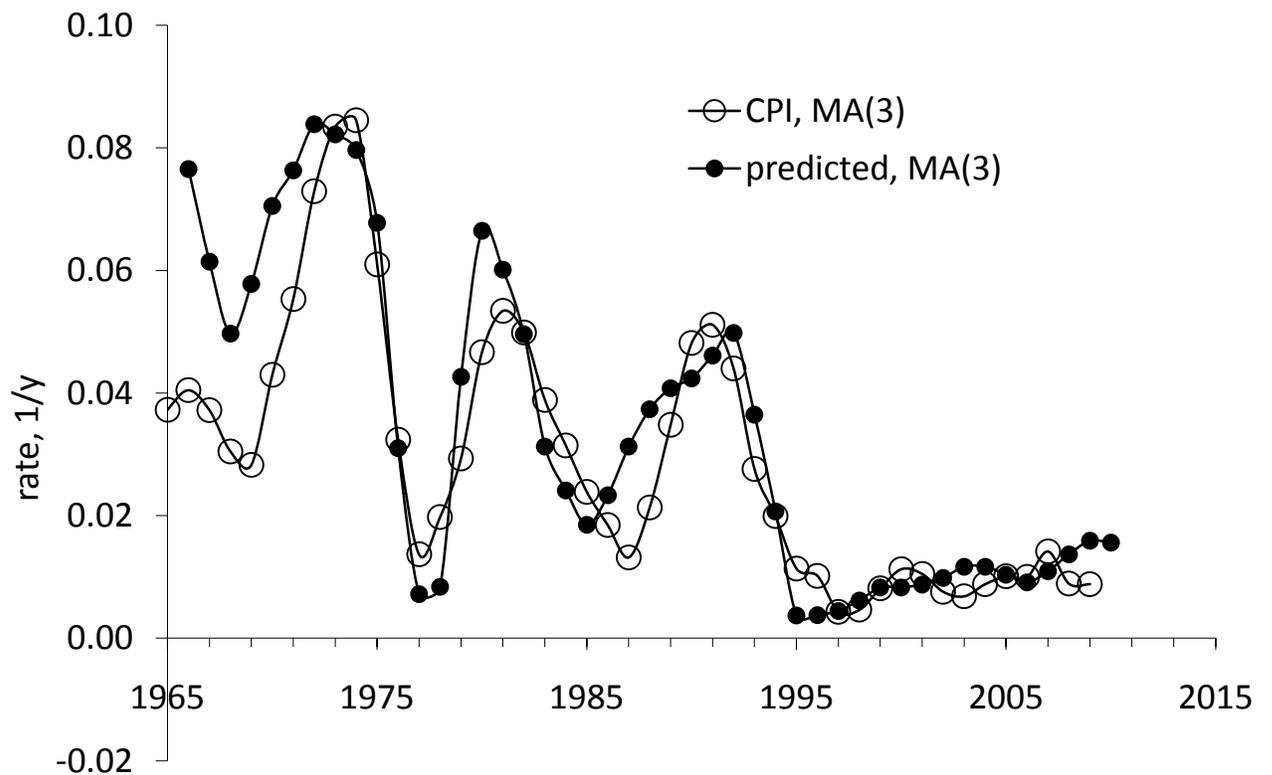

**Figure 7: The observed and predicted rate of CPI inflation in Figure 6 smoothed with MA(3). The overall agreement is very good.**

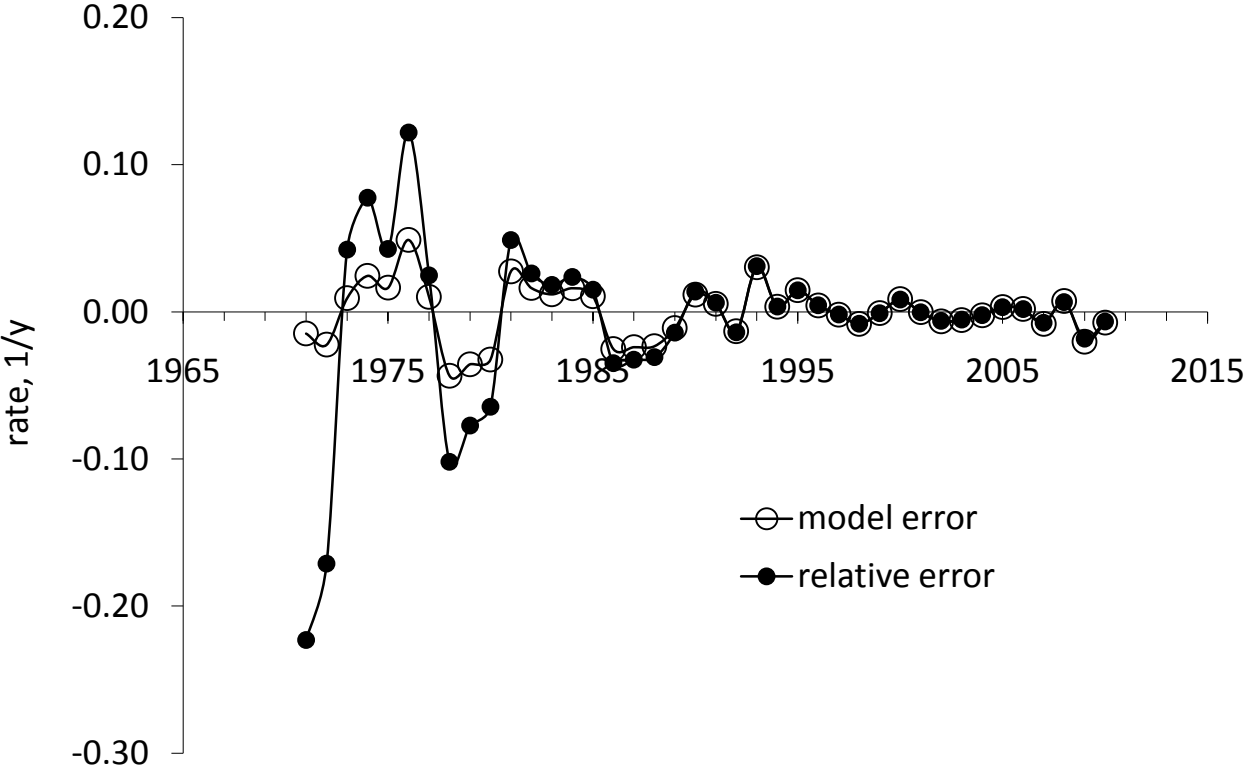

**Figure 8: The model error and relative model error.**

The observed CPI time series between 1967 and 2010 contains 44 readings. It is a small sample for statistical inferences. Nevertheless, we have conducted the Engle-Granger and Johansen tests for cointegration. Tables 3 and 4 list some principal results. At first, we test for unit roots the model residual error. All tests in Table 3 reject the null of a unit root. The only difference with the authentic Engle-Granger approach is that we have used the residual obtained by visual fit of the cumulative curves in Figure 6 instead of a linear regression. Our previous experience shows that linear regression is hardly the best way to estimate coefficients in (7). In any case, Table 3 implies that model (7) represent a long term equilibrium relation between inflation and the change in labour force. In the absence of AR-terms and the one-year lead of the dFL/LF, this relation is a causal one.

**Table 3: Unit root tests of the model residual**

| DF | -5.51* | -3.63 |
|---|---|---|
| **DF-GLS [lag]** | | |
| 4 | -3.1 | -3.77 |
| 3 | -2.804 | -3.77 |
| 2 | -4.318* | -3.77 |
| 1 | -6.97* | -3.77 |
| **PP** | | |
| z(ρ) | -31.91* | -18.43 |
| z(t) | -5.43* | -3.63 |

The Johansen test for cointegration results in rank = 1, i.e. one cointegrating relation between the observed and predicted time series, for both trend specifications: "*none*" and "*rconstant*". The largest lag in these tests is 2 years. Considering the principal results of both cointegration tests we can conclude that (CPI) inflation and labour force are cointegrated between 1967 and 2010. This validates the results of the estimated linear regression and other statistical inferences.

We have already estimated $R^2 = 0.64$ for the annual time series. Corresponding RMFSE, i.e. RMS value of the model error, is of $0.018y^{-1}$. This is the accuracy of the CPI inflation prediction at a one year horizon. As expected, a VAR model with maximum lag of 2 years, i.e. the inclusion of corresponding autoregressive terms in (7), provides only a marginal improvement on the deterministic model: $R^2 = 0.74$ and RMSFE = $0.013y^{-1}$.

**Table 4: The Johansen rank test, maxlag = 2**

| none | rank | LL | eigenvalue | trace stat | 5% critical |
|---|---|---|---|---|---|
| | 0 | 220.6 | . | 34.7 | 12.53 |
| | 1 | 236.4 | 0.53 | 3.20* | 3.84 |
| | 2 | 238 | 0.07 | | |
| rconstant | rank | LL | eigenvalue | trace stat | 5% critical |
| | 0 | 220.6 | . | 39.95 | 12.53 |
| | 1 | 236.4 | 0.53 | 8.44* | 9.42 |
| | 2 | 238 | 0.18 | | |

A VEC model with maximum lag 2 years and one cointegrating relation, results in a larger RMSFE of $0.014y^{-1}$. Statistically, a much better result is obtained when MA(3) of the predicted series is used. This is a correct operation since only past readings of the LF are used and the prediction is effectively an out-of-sample one. For the period between 1971 and 2010 we have obtained $R^2 = 0.80$ and RMSFE = $0.011y^{-1}$, with the naive RMSFE of $0.016y^{-1}$.

The goodness-of-fit for the cumulative curves is indistinguishable from 1.0. Therefore, it could be argued that measurements of the CPI in the long run are redundant and can be completely replaceable by the level of labour force, which is obviously much easier to estimate with a very high accuracy than inflation. Apparently, CPI is not an unambiguous macroeconomic variable and includes subjective judgments on hedonic pricing to correct for quality changes or when new goods and services are introduced.

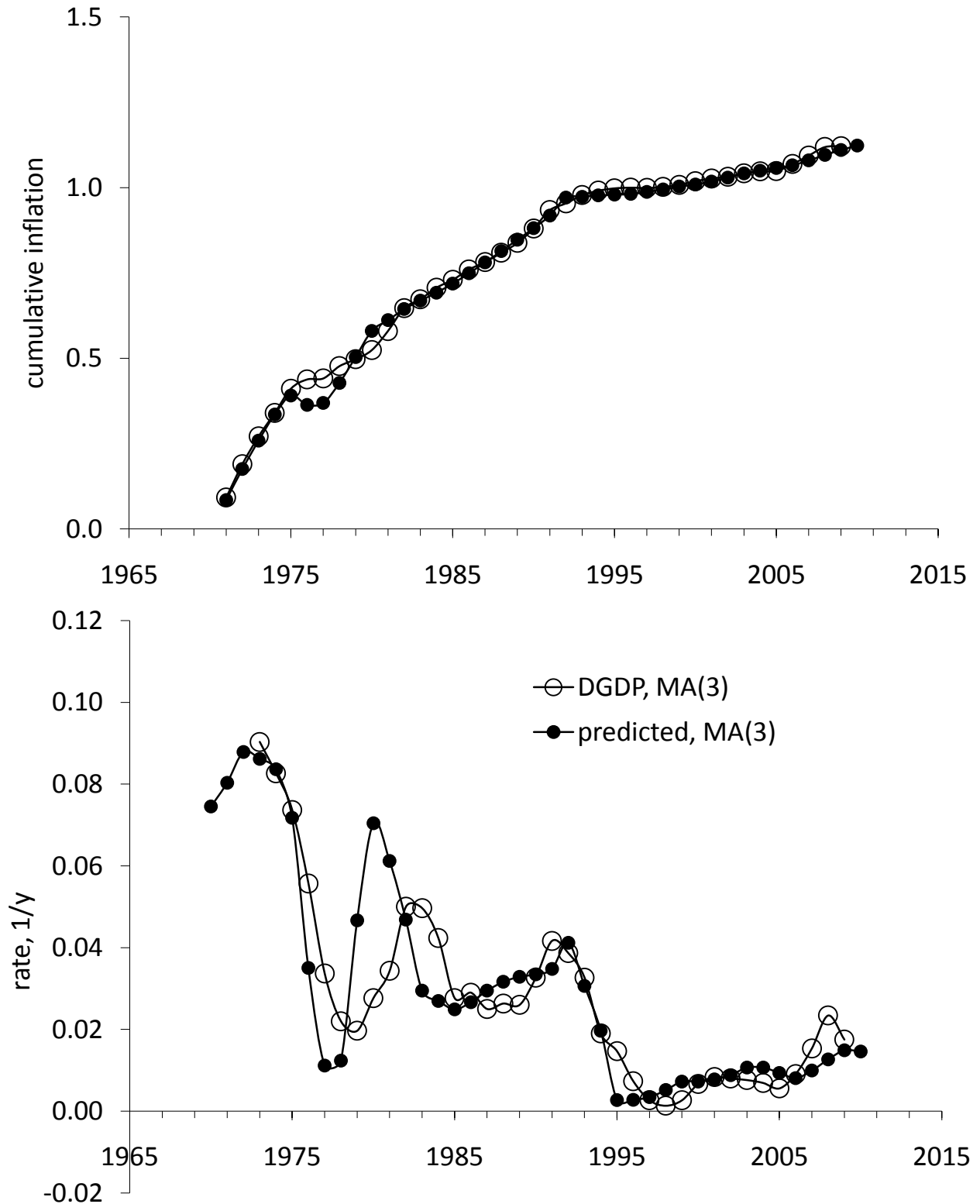

**Figure 9:** *Upper panel*: **Cumulative curves of the observed and modelled DGDP between 1971 and 2009.** *Lower panel*: **The annual DGDP smoothed with MA(3) since 1973.**

Now, we turn our attention to the second measure of inflation, the GDP deflator. The annual DGDP time series starts in 1971. The period between 1971 and 2009 is split into three segments and the (visual) best-fit linear lagged relationships are as follows:

$$DGDP(t) = 1.9dLF(t-2)LF(t-2) + 0.057; t < 1980$$
$$DGDP(t) = 0.5dLF(t-2)/LF(t-2) + 0.021; 1992 > t > 1980$$
$$DGDP(t) = 0.5dLF(t-2)/LF(t-2) + 0.006; t > 1991 \qquad (8)$$

In (8) coefficients are slightly different from those in (7). The most prominent is the change in slope from +1.3 to +0.5 between 1980 and 1991. This effect is associated with low resolution for this particular period when dLF/LF was almost constant with a small trough around 1983. The change in slope can be replaced by a proportional change in free term - from 0.008 to 0.021. The curvature of the observed DGDP curve between 1980 and 1992 also adds to the change in the slope. Otherwise, both models are very similar and identical since 1992.

Figure 9 depicts the observed and predicted DGDP curves - cumulative and MA(3) ones. The length of the DGDP series is only 38 readings and all statistical inferences would be unreliable. For this reason, we skip cointegration tests and linear regression analysis. On the other hand, the CPI and DGDP curves in Figure 1 are very close, and thus, statistical results should be similar.

$$DGDP(t) = 1.9dLF(t-2)LF(t-2) + 0.057; t < 1980$$
$$DGDP(t) = 0.5dLF(t-2)/LF(t-2) + 0.021; 1992 > t > 1980$$
$$DGDP(t) = 0.5dLF(t-2)/LF(t-2) + 0.006; t > 1991 \qquad (9)$$

The CPI annual series has 44 readings between 1967 and 2010. Small samples cannot provide robust statistical estimates and inferences. The OECD reports quarterly estimates of inflation and labour force that are however both noisy because of the significant measurement errors. As we use the cumulative curve approach these measurement errors have to be suppressed in the long run by destructive interference. One can estimate relevant coefficients in (8) for the quarterly readings of the DGDP and dLF/LF.

Figure 10 presents both the quarterly and cumulative curves of the observed and predicted inflation rate. The quarterly measured inflation is smoothed with MA(4) and the predicted one with MA(8) because the original Q/Q growth rates are highly volatile. The agreement between the cumulative curves is again excellent.

5. **The generalized model**

We have estimated several individual links between labour force, unemployment and inflation. The change in labour force and the rate of inflation are cointegrated, as the Engle-Granger and Johansen tests have shown. For the rate of unemployment, the model has no sense because of low resolution. In this situation, the generalized model is somewhat obsolete. However, we have estimated this model for methodological purposes as well as for the completeness of our concept. We split the entire modelling period into two segments and obtain the following relationship:

$$CPI(t) = 1.8dLF(t-1)LF(t-1) - 6.0UE(t-1) + 0.04; t < 1989$$
$$CPI(t) = 0.4dLF(t-1)LF(t-1) - 0.7UE(t-1) + 0.03; t > 1988 \qquad (10)$$

Figure 11 presents the measured and predicted CPI inflation. As discussed above, the CPI does not represent the economy as a whole, and thus, the CPI evolution is not necessarily a one-to-one reaction to the change in labour force.

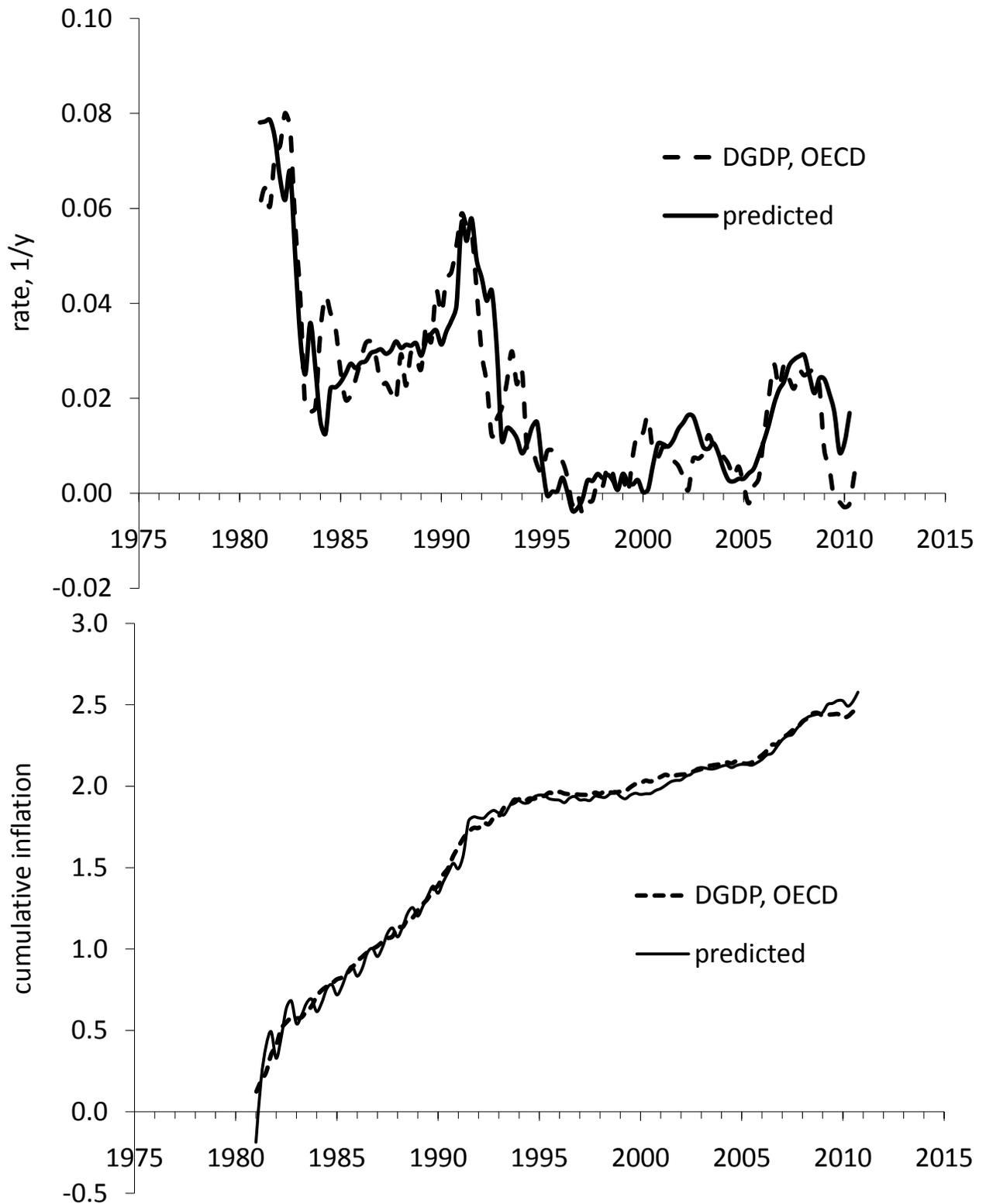

**Figure 10:** *Upper panel*: Quarterly readings of the observed and modelled rate of the overall price inflation as represented by the GDP deflator, DGDP. According to the changes in labour force definition in 1992, the predicted curve is a piecewise linear and lagged function. *Lower panel*: Cumulative curves of those in the upper panel since 1981.

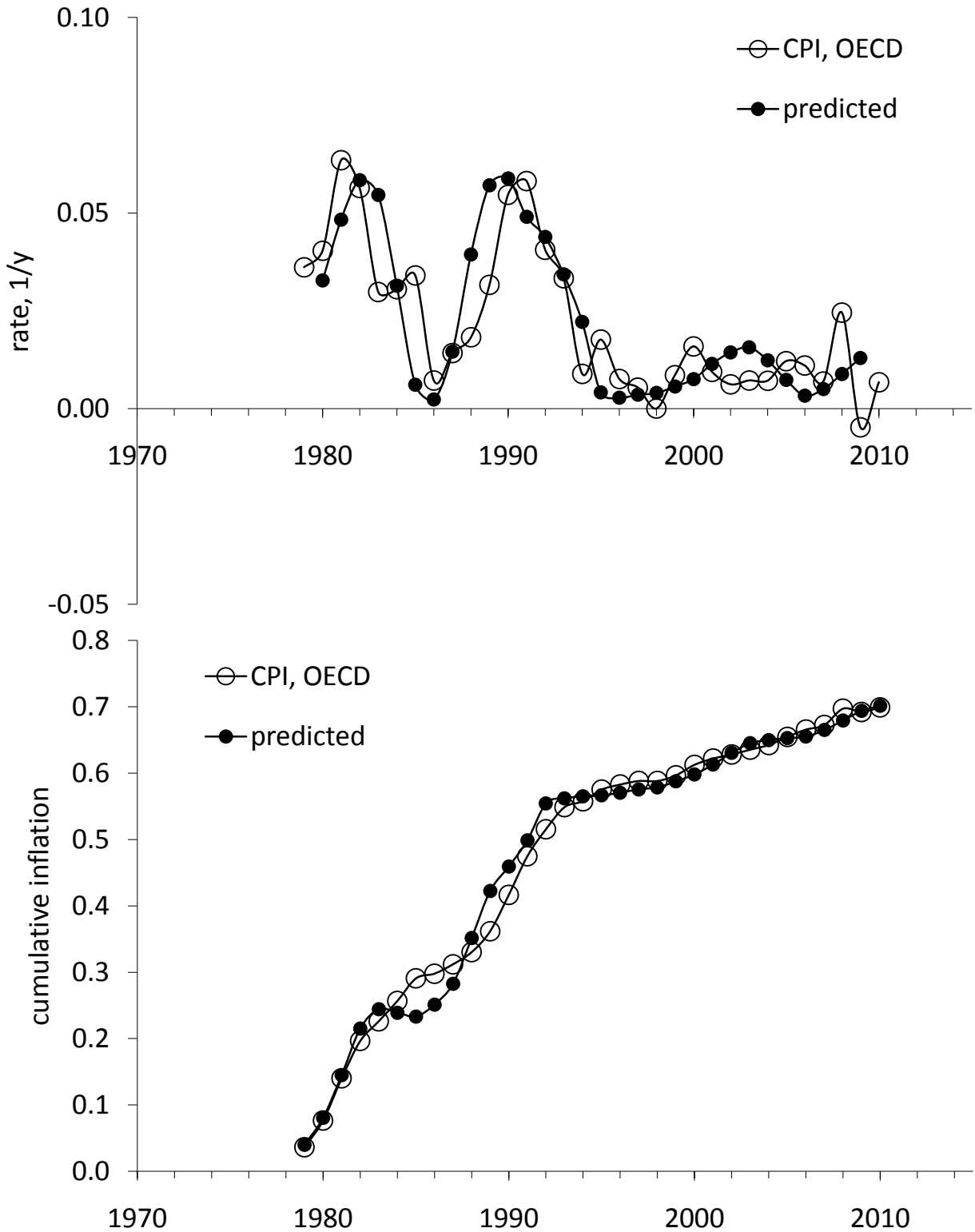

**Figure 11:** *Upper panel:* **The rate of the overall price inflation, CPI, in Switzerland as predicted by the generalized model with a structural break neat 1993 related to the change in measuring units. Notice that the predicted series is smoothed with MA(3).** *Lower panel:* **The relevant cumulative curves.**

**Conclusion**

The rate of price inflation and unemployment in Switzerland is a one-to-one correspondence to the change in labour force. This conclusion validates earlier models for many developed countries: the U.S., Japan, Germany, France, Italy, Canada, the Netherlands, Sweden, Austria, and Australia. Switzerland was not studied in detail so far. The excellence of the obtained statistical and conceptual results compensates the delay in analysis.

Overall, we have established that there exist long term equilibrium relations the rate of labour force change and the rate of inflation. In this sense, money factor plays no significant role in the low inflation level observed in Switzerland since the 1990s, as suggested by Reynard (2006). The level of statistical significance of these cointegrating relations allows us to consider these links as deterministic ones, as adopted in physics. Unlike the New Keynesian Phillips curve models, the relationships proposed in this paper do not use autoregressive properties of any macroeconomic variable under consideration.

The change in labour force includes a strong demographics component and thus is stochastic to the extent the evolution of population in a given country is stochastic. Since the level of labour force is a measurable value one does not need to estimate its stochastic properties - they are obtained automatically with tedious routine measurements. The Swiss Federal Statistics Office (2011) provides several long-term projections of the labour force level. We have used three of these projections with high, low and middle fertility scenarios. Figure 12 depicts three predictions of inflation based on these projections and the most recent linear relationship between dLF=LF and CPI. In all cases, the rate of inflation will be below 1% per year in the next 40 years. There is no long term danger of deflation, however. Smooth and low price inflation in the decades to come will likely make aggressive reaction of the SNB, as based on the Taylor rule (Perruchoud, 2009), less popular. Based on relationship (5), we expect the rate of unemployment in Switzerland to be slightly above 4.0% over the next four decades.

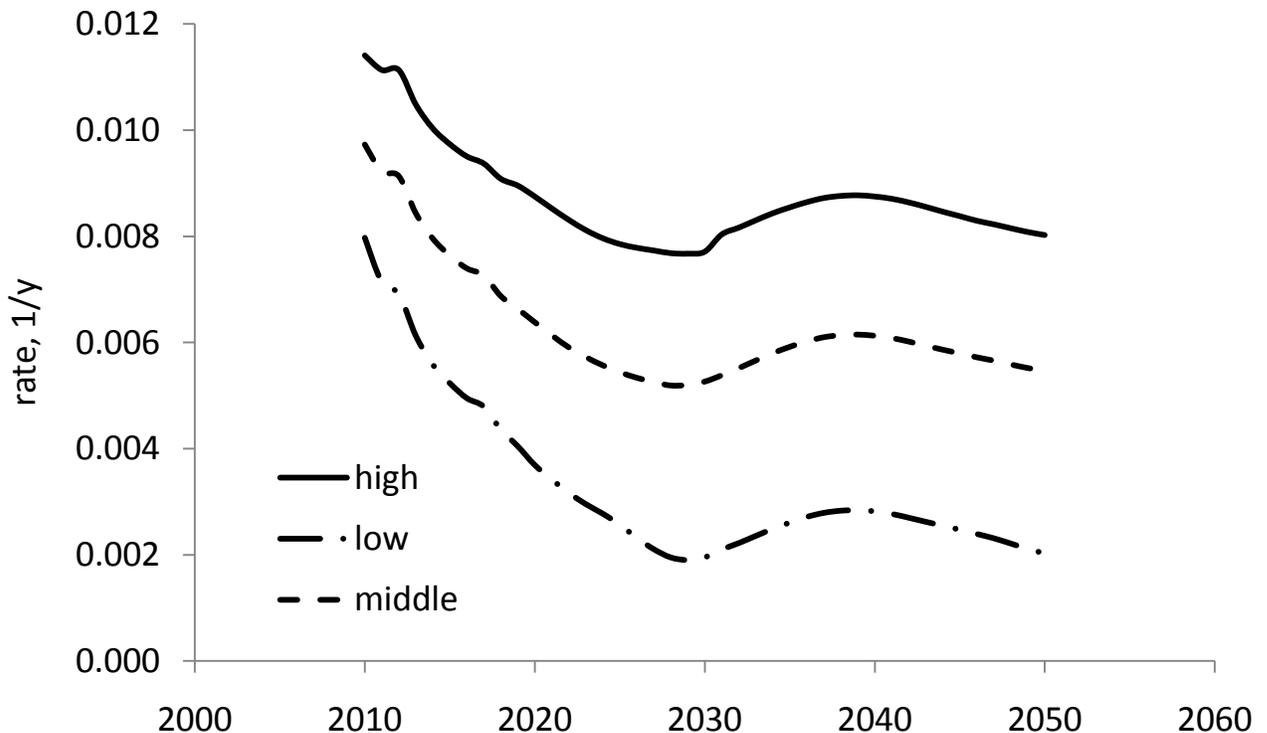

**Figure 12: The prediction of CPI inflation between 2010 and 2050.**